# THE THERMAL REGULATION OF GRAVITATIONAL INSTABILITIES IN PROTOPLANETARY DISKS. IV. SIMULATIONS WITH ENVELOPE IRRADIATION

Kai Cai

Department of Physics and Astronomy, McMaster University, Hamilton, ON, Canada L8S 4M1

Richard H. Durisen, Aaron C. Boley

Department of Astronomy, Indiana University, 727 E. 3$^{rd}$ St., Bloomington, IN 47405

Megan K. Pickett

Department of Physics, Lawrence University, Box 599, Appleton, WI 54911

and

Annie C. Mejía

Museum of Flight, Exhibits Department, 9404 E. Marginal Way South, Seattle, WA 98108

# ABSTRACT

It is generally thought that protoplanetary disks embedded in envelopes are more massive and thus more susceptible to gravitational instabilities (GIs) than exposed disks. We present three-dimensional radiative hydrodynamics simulations of protoplanetary disks with the presence of envelope irradiation. For a disk with a radius of 40 AU and a mass of 0.07 $M_\odot$ around a young star of 0.5 $M_\odot$, envelope irradiation tends to weaken and even suppress GIs as the irradiating flux is increased. The global mass transport induced by GIs is dominated by lower-order modes, and irradiation preferentially suppresses higher-order modes. As a result, gravitational torques and mass inflow rates are actually increased by mild irradiation. None of the simulations produce dense clumps or rapid cooling by convection, arguing against direct formation of giant planets by disk instability, at least in irradiated disks. However, dense gas rings and radial mass concentrations are produced, and these might be conducive to accelerated planetary core formation. Preliminary results from a simulation of a massive embedded disk with physical characteristics similar to one of the disks in the embedded source L1551 IRS5 indicate a long radiative cooling time and no fragmentation. The GIs in this disk are dominated by global two and three-armed modes.

*Subject headings:* accretion, accretion disks – hydrodynamics – instabilities – planetary systems: formation – planetary systems: protoplanetary disks

# 1. MOTIVATION

By the early 1990s, it was clear that infalling dusty envelopes must be invoked to explain the large infrared (IR) excesses present in the observed spectral energy distributions (SEDs) of Class I YSOs and most of the "flat-spectrum" T Tauri stars (Kenyon et al. 1993a, Kenyon et al. 1993b, Calvet et al. 1994; however, see Chiang et al. 2001 for a different view). D'Alessio et al. (1997) explored the effects of irradiation from infalling optically thick protostellar envelopes on a standard viscous accretion disk and found that, although the outer disk will be heated significantly, it may still be unstable against gravitational instabilities (GIs), at least in the case of HL Tau.  In contrast, the outer disk is generally stabilized against GIs by the more intense irradiation from the central star (D'Alessio et al. 1998).

Over the past few years, millimeter surveys of embedded sources (e.g. Looney et al. 2000, 2003, Motte & André 2001) have shown evidence for the presence of huge envelopes and possibly massive disks (Eisner & Carpenter 2006). Although the disk masses are still somewhat controversial (Looney et al. 2003), the total circumstellar masses are undoubtedly larger than those of YSOs at later stages (see also Eisner & Carpenter 2003). The larger circumstellar disk mass and the smaller disk size at early times (e.g. Durisen et al. 1989, Stahler et al. 1994, Pickett et al. 1997) work toward enhancing the surface density of the disk, thus lowering the Toomre $Q$-value (Toomre 1964) and making the disk more susceptible to the development of GIs. As an example, Osorio et al. (2003) studied the L1551 IRS5 binary system in unprecedented detail and estimated disk parameters based on fitting the observed fluxes with irradiated steady-state $\alpha$-disk models. They found that the outer part of the disks have $Q < 1$.

All this work suggests that GIs are more likely to occur in protoplanetary disks during the embedded phase. Fragmentation of disks due to GIs has been proposed as a formation mechanism for gas giant planets (Boss 1997, 1998, 2000, Mayer et al. 2002, 2004). The strength of GIs depends critically on the thermal physics of the disk (see Durisen et al. 2007 for a review). In particular, efficient cooling is required for the disk to fragment into clumps, and so it is important to determine how strong GIs can be under realistic thermal conditions, including envelope irradiation.

Previous disk GI simulations by our group focused on naked disks, which would correspond to Class II or later sources (Pickett et al. 2003, hereafter Paper I; Mejía et al. 2005, hereafter Paper II; Boley et al. 2006, hereafter Paper III). Boss (2001, 2002, 2003, 2005, 2006), in his disk simulations with radiative diffusion, mimics the effect of envelope irradiation by embedding the disk in a constant temperature thermal bath. Dense protoplanetary clumps form in his simulations, and he attributes this to rapid cooling by convection (Boss 2004). Employing a similar treatment to that of Boss in the optically thick region but using different atmospheric boundary conditions (BCs), Mejía (2004) and Paper III found that, without envelope irradiation, no persistent clumps form and convection does not cause rapid cooling in a GI-active disk. This drastic difference in outcome suggests that results are sensitive to radiative BCs.

Therefore, from both observational and theoretical perspectives, it is interesting to investigate the strength of GI activity when the disk experiences IR irradiation from a surrounding envelope. This is directly applicable to GIs in embedded disks and permits more direct comparisons with Boss simulations. Preliminary work by Mejía (2004) on a *stellar* irradiated disk indicates that GIs are weakened. However, stellar irradiation is difficult to model on coarse 3D grids due to the high optical depth of disks to starlight (Mejía 2004). For this reason, we focus our efforts here on treating the simpler case of envelope IR irradiation.

This paper is the fourth in our series on the regulation of GIs in protoplanetary disks by thermal processes. Papers I and II featured disk simulations with idealized cooling prescriptions. Paper III presented our first simulation using the radiative cooling scheme developed in Mejía (2004). In this paper, we present results of disk simulations using a version of Mejía (2004)'s routines modified to include envelope irradiation. This new method, described in §2, has already been used to study the effects on GIs of varying metallicities and dust grain sizes in Cai et al. (2006). The set of simulations and their results are presented in §3. Section 4 discusses the significance of the results and compares them to work by other groups, and §5 is a brief summary of our conclusions.

# 2. METHODS

## 2.1 3D Hydrodynamics

We conduct protoplanetary disk simulations using the Indiana University Hydrodynamics Group code (Pickett et al. 1998, 2000, Mejía 2004, Papers I & II), which solves the equations of hydrodynamics in conservative form on a cylindrical grid $(r, \varphi, z)$ to second order in space and time using an ideal gas equation of state with the ratio of specific heats $\gamma = 5/3$. Self-gravity and shock mediation by artificial bulk viscosity are included. Reflection symmetry is assumed about the equatorial plane, and free outflow boundaries are used along the top, outer, and inner edges of the grid.

## 2.2 Radiative Scheme

Let $\tau$ be the Rosseland mean optical depth measured vertically down from above. As in Mejía (2004) and Paper III, the cells with $\tau \geq 2/3$ are considered part of the disk interior, and the energy flow in all three directions is calculated using flux-limited diffusion (Bodenheimer et al. 1990). The cells with $\tau < 2/3$ (the atmosphere) cool according to a free-streaming approximation, so they cool as much as their emissivity allows. However, similar to the original Mejía (2004) scheme, the atmospheric cells directly above the disk interior are also heated by the flux from underneath, which is determined by the boundary condition described below. This emergent flux is attenuated as it passes upward through the atmosphere according to $\exp(-\tau_{up})$, where $\tau_{up}$ is the Planck mean optical depth, measured between the cell of interest and the interior/atmosphere boundary, evaluated at the local temperature. The Planck mean is used in this regime because the medium is optically thin.

For the simulations presented in this paper, we include external infrared irradiation from an envelope heated by the star. This IR irradiation is approximately treated as a blackbody flux $\sigma T_{irr}^4$ with a characteristic temperature $T_{irr}$ shining vertically downward onto the disk. For one of the calculations, as noted below in §3.2, we include a similar envelope heating flux entering radially from the outer boundary of the disk, because a real envelope is usually much larger than the disk. Also, for this reason, an envelope irradiation with constant temperature (i.e., neglecting the drop off of envelope temperature farther out from the central star) may not be a bad approximation (e.g., the disks in L1551 IRS 5, D'Alessio 2005, private communication). In each column of the atmosphere, the downward envelope flux is reduced by a factor 1-exp(-$\Delta\tau_P$) as it crosses each cell, where $\tau_P$ is the Planck mean optical depth measured vertically downward and evaluated at $T_{irr}$ and where $\Delta\tau_P$ is the change in $\tau_P$ across a cell. Unlike Boss (2003), we do not explicitly model an infalling envelope here, nor does the disk atmosphere include the infall effects.

The total cooling rate in atmosphere cells can then be expressed as

$$\Lambda = \rho\kappa_{Planck}(T)(4\sigma T^4 - \sigma T_{int}^4 e^{-\tau_{int}}) - \rho\kappa_{Planck}(T_{irr})\sigma T_{irr}^4 e^{-\tau_P},$$ 

(1)

where heating by the upward boundary flux leaving the hot disk interior and heating by the downward envelope irradiation flux are treated as additional terms to the free-streaming approximation. Note that all terms use the Planck mean opacity $\kappa_{Planck}$, but it is evaluated at different temperatures in different terms. The term $(-\rho\kappa_{Planck}(T)\sigma T_{int}^4 e^{-\tau_{int}})$ represents heating by the flux leaving the interior and is discussed in more detail below. A three-cell linear smoothing of $\Lambda$ in the radial direction is also employed to make the temperature structure smoother.

As in Mejía (2004) and Paper III, both the disk interior and the atmosphere are treated separately and then coupled by an Eddington-like atmosphere boundary condition, which defines the boundary flux for the diffusion approximation. This boundary flux represents the balance between the energy leaving the disk interior through the atmosphere and the energy gained by the atmosphere from external heating sources that must be carried down to the interior. Because there is an additional source of heating, i.e., envelope irradiation, the boundary condition defined by equation (2-26) in Mejía (2004) must be modified to incorporate the contribution of the (residual) envelope irradiation flux that survives atmospheric absorption:

$$F_{bdry} = \frac{4\,\sigma\,(T_{bdry}^4 - T_{atm}^4 - T_{irr}^4 e^{-\tau_P})}{3\,(\tau_{bdry} + 2/3)}.$$
(2)

where $T_{bdry}$ and $\tau_{bdry}$ are taken to be the values of the *first* optically thick cell right below the boundary, as explained by Mejía (2004). The atmospheric temperature $T_{atm}$ is a measure of the downward flux onto the interior of the disk due to emission by the atmosphere. It is assumed to be a half of the total atmospheric cooling. After the boundary flux $F_{bdry}$ is computed, a photospheric temperature can be determined:

$$T_{ph} = (F_{bdry}/\sigma + T_{atm}^4 + T_{irr}^4 e^{-\tau})^{1/4} \qquad . \quad (3)$$

It is then used in (1). When $\tau_{bdry} = 2/3$ (exactly), equations (2) and (3) indicate $T_{ph}$ reduces to $T_{bdry}$, making the scheme self-consistent.

In Mejía's original scheme, there is no photospheric boundary cell. Every cell with $\tau \geq 2/3$ is an interior cell, and every cell with $\tau < 2/3$ is an atmosphere cell. In some columns of the disk, especially the inner disk, the abrupt change in $\tau$ from the *first* optically thick cell to its optically thin neighbor above can be so large that the heat from the hot interior underneath creates an over-heated layer above the boundary. For the simulations in this paper, a one-cell transition layer in the vertical direction consisting of these first optically thin cells above the interior is introduced to eliminate this numerical problem. In any column, the temperature of the transition cell is determined by the Eddington-like fit in that column (eq. [2]), and transition cells are otherwise neither heated nor cooled.

Boley et al. (2007c; see also Paper III) describe a test suite for radiative hydrodynamics algorithms that are used to evolve protoplanetary disks. One of these tests assumes a functional form for an *ad hoc* energy dissipation in a vertically stratified gas layer under constant gravity and reflection symmetry about a "midplane". For this test, code results can be compared to analytic solutions. Figure 1 shows the flux and temperature profiles that Mejía (2004)'s radiative scheme achieves in a case without irradiation where numerical oscillations occur because $\tau = 2/3$ is too close to a cell boundary. Also shown in Figure 1 is the result for the modified scheme with a one-cell transition layer. The new scheme not only yields a more realistic vertical temperature structure, but as discussed in more detail by Boley et al. (2007c), it also suppresses the numerical oscillations. The sudden drop in the temperature in both schemes at the photosphere (cell 11 or 12) is a result of the lack of complete cell-to-cell coupling in the optically thin region. For both cases, accurate fluxes and temperature structures are reproduced throughout the disk interior.

Following Mejía (2004) and Paper III, the volumetric radiative cooling rate $\Lambda$, the volumetric shock heating rate due to artificial viscosity $\Gamma$, and the divergence of the radiative flux $\nabla \cdot \boldsymbol{F}$ are limited such that the cooling and heating times are not allowed to be shorter than about 3% of the initial orbital time of the outer disk. As discussed in Paper III, these limiters do not prevent fragmentation, and the cells that are affected tend to be located in the diffuse background, not within areas of interest.

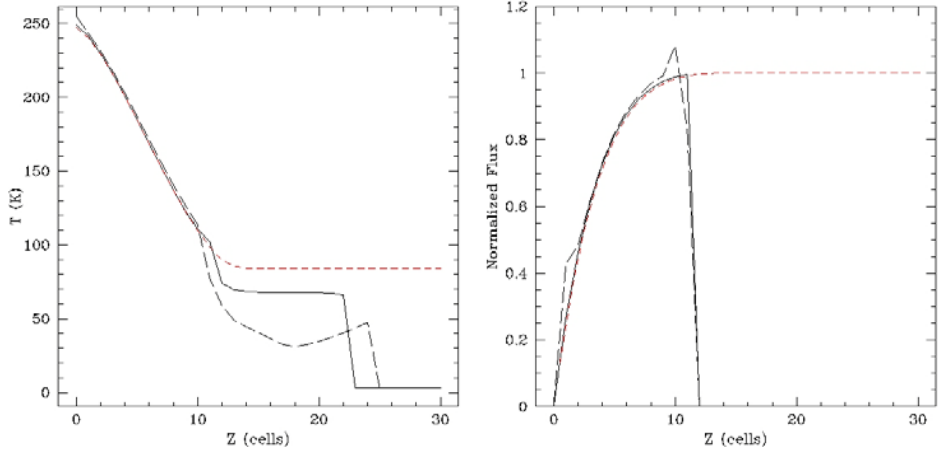

**Figure 1.** Snap shot results of relaxation tests described in Boley et al. (2007c). The left panel indicates the analytic temperature profile (dashed, red [gray] curve), the actual profile for the Mejía (2004) scheme (dashed, black curve), and the actual profile for the one-cell transition layer modification (solid, black curve). The first drop indicates where the disk interior ends and the atmosphere begins, and is due to the lack of complete cell-to-cell coupling that the solution to the equation of radiative transfer requires (see discussion in Paper III). The second drop is where the density falls to background. The right panel indicates the flux through the interior only, because the flux through the atmosphere is not explicitly tracked. The addition of the transition layer gives more accurate temperature and flux distributions and also avoids the pulsations that occur in the Mejía (2004) scheme when the photosphere is near a vertical grid boundary.

## 2.3 Boss-like Boundary Condition

As mentioned in §1, we suspect that the apparent disagreement in cooling rates between our results and those of Boss is primarily due to differences in the treatment of radiative boundary conditions (BCs), although differences in the treatment of the equation of state may also be important (Boley et al. 2007a). In order to investigate the effects of different BCs on the same radiative cooling algorithm, we implement a Boss-like BC for our flux-limited diffusion scheme in some comparison calculations discussed in §3.4. For regions where $\tau \geq 10$, we compute radiative diffusion in all three directions with the flux limiter turned off, as done in Boss (2001). In the outer disk (the optically thin region) where $\tau < 10$, the temperature is set to the background temperature $T_B$. The two regions are not otherwise coupled radiatively. Our initial model, grid geometry, and internal energy for molecular hydrogen are still different from those of Boss.

## 3. THE SIMULATIONS

### 3.1 Initial Model for the Irradiation Study

For purposes of comparison with Paper III, the same initial axisymmetric equilibrium disk is used for the calculations in which $T_{irr}$ is varied. The disk is nearly Keplerian, has a mass 0.07 $M_{\odot}$ with a surface density $\Sigma(r) \propto r^{-0.5}$ from 2.3 to 40 AU, and orbits a star of 0.5 $M_{\odot}$. The initial minimum value of the Toomre stability parameter $Q_{min}$ is about 1.5, so the disk is marginally unstable to GIs. More details can be found in Paper III and in Mejía (2004).

### 3.2 The Simulations With Varied Envelope Irradiation

We present three new simulations, each with a different $T_{irr}$. The simulation with $T_{irr} = 15$K, hereafter Irr 15K, is the same as the solar metallicity simulation (Z = $Z_{\odot}$) presented by Cai et al. (2006). The second simulation (Irr 25K) was run with $T_{irr} = 25$K. The third simulation with $T_{irr} = 50$K (Irr 50K), which demonstrates the damping of GIs by strong envelope irradiation, is started at time $t = 11.5$ ORP of Irr 25K run. Here the time unit of 1 ORP is defined as the initial outer rotation period at 33 AU, about 253 yr. For comparison, we include the non-irradiated case (No-Irr) from Paper III. Except for No-Irr, the initial grid has (256, 128, 64) cells above the midplane in cylindrical coordinates ($r, \varphi, z$). The No-Irr run requires only 32 vertical cells because the atmosphere of the non-irradiated disk is less expansive. The number of radial cells is increased from 256 to 512 when the disk expands radially due to the onset of GIs. In the Irr 15K case only, we include an extra envelope irradiation shining horizontally inward from the grid outer boundary (see §3.2.2). Each simulation is evolved for solar metallicity with a maximum grain size for the D'Alessio et al. (2001) opacities of $a_{max} = 1$ $\mu$m and a mean molecular weight of about 2.7 (see Paper III).

Envelope temperatures of 15K and 25K are at the low end of likely envelope temperatures (Chick & Cassen 1997). They are probably consistent with the low mass of our star and are lower than the 50K bath used in Boss's (2001, 2002) simulations. Boss's disks and our disks have similar masses, but his have an initial outer radius of 20AU, versus 40AU for our disks. The lower surface density for our disks means a lower $T_{irr}$ is required to achieve a marginally unstable $Q$-value.

One way to interpret $T_{irr}$ is to ask what would be the radius of a spherical envelope shell radiating down on the disk like a $T_{irr}$ blackbody if the shell is maintained through the absorption of $L_{star}$. We then have $L_{star} = 4\pi R_{shell}^{2} \sigma T_{irr}^{4}$ and $R_{shell} = R_{star}(T_{star}/T_{irr})^{2}$. The parameters of the protostar are chosen to be the same as in D'Alessio et al. (2001) and Mejía (2004), where $T_{eff} = 4000$K and $R_{star} = 2$R$_{\odot}$, values typical of T Tauri stars (see, e.g., Chiang & Goldreich 1997, 2001). For $T_{irr} = 15$K, $R_{shell} = 661$ AU, which is a reasonable size for a protostellar envelope (e.g., Osorio et al. 2003). For $T_{irr}$ = 25K and 50K, $R_{shell}$ = 238 AU and 59.5 AU, respectively.

It is also interesting to compare the energy input onto the disk with the stellar luminosity. Assuming the disk is flat, the ratio between the total luminosity of envelope irradiation that shines down onto the disk $L_{env}$, and the stellar luminosity $L_{star}$ is then given by

$$\frac{L_{env}}{L_{star}} = \frac{1}{2}\left(\frac{T_{env}}{T_{star}}\right)^4\left(\frac{R_d}{R_{env}}\right)^2 = 2.25\times10^{-11}T_{irr}\,(K)^4 R_d\,(AU)^2 \qquad (4)$$

With $T_{irr}$ = 15K and a disk radius $R_d$ = 50 AU (see §3.2.2 for an explanation), $L_{env}$ amounts to about $2.8\times10^{-3}$ of the stellar luminosity. For $T_{irr}$ = 25K and the same disk radius, $L_{env}/L_{star}$ = $2.2\times10^{-2}$. When $T_{irr}$ is raised to 50K, the envelope irradiation becomes approximately 1/3 of the stellar luminosity. Please note, however, these are only simple analytic calculations and are not what we put into the disk. A tally of the actual energy input in the disk is presented in §3.2.2.

**Table 1** Characteristics of the Disk Models

| Case | No-Irr | Irr 15K | Irr 25K | Irr 50K |
|---|---|---|---|---|
| $T_{env}$ | 0 | 15K | 25K | 50K |
| Duration[a] | 16.0 | 15.7 | 15.4 | 4.4 [b] |
| $t_1$ (ORPs) | 3.0 | 5.0 | 7.0 | NA |
| $t_2$ (ORPs) | 10.0 | 9.9 | 9.9 | NA |
| $\langle A \rangle$ | 1.51 | 1.16 | 1.01 | 0.40 |
| $\langle Q_{late} \rangle$ | 1.51 | 1.65 | 1.96 | 2.64 |
| $Q_{min}$ | 1.28 | 1.32 | 1.44 | 2.20 |
| $\Delta r_{Q<2}$ (AU) | 33.3 | 25.7 | 13.3 | 0 |
| $t_{cool}$[a] | 2.7 | 3.2 | 5 | 9 |
| $L_{env}$(code) (erg/s) | 0 | $8.6 \times 10^{30}$ | $7.2 \times 10^{31}$ | $1.3 \times 10^{33}$ |
| analytic $L_{env}$($R_d$ = 50AU) (erg/s) | 0 | $1.0 \times 10^{31}$ | $7.8 \times 10^{31}$ | $1.3 \times 10^{33}$ |
| $R_{shell}$ (AU) | | 661 | 238 | 59.5 |
| $L_{disk}$ (erg/s) | $4.7 \times 10^{30}$ | $5.3 \times 10^{30}$ | $5.0 \times 10^{30}$ | $4.7 \times 10^{30}$ |
| Average $\dot{M}$ ($M_\odot$ yr$^{-1}$) | $2 \times 10^{-7}$ | $6.8 \times 10^{-7}$ | $5.0 \times 10^{-7}$ | NA |

[a] All times are given in units of an ORP.
[b] From 11.5 to 15.9 (ORPs).

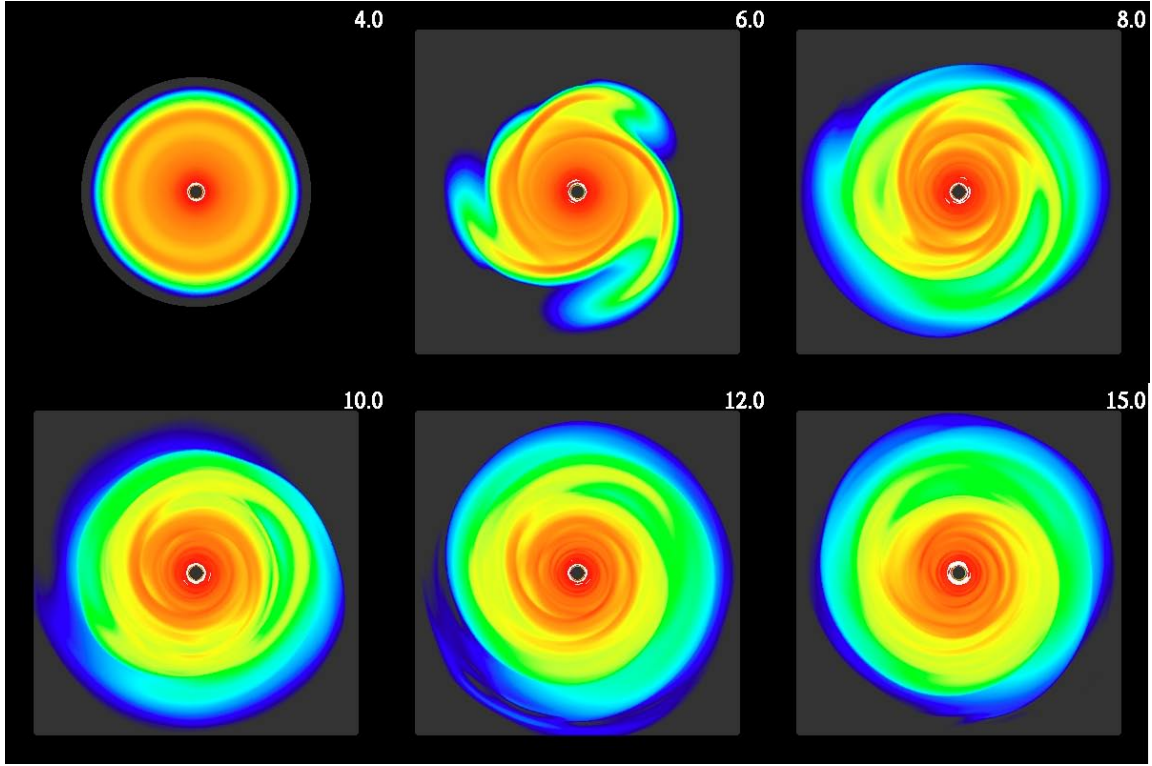

**Figure 2.** Midplane density maps of Irr 15K showing the evolution. Each square enclosing the disk is 121.4 AU on a side. Densities are displayed on a logarithmic color scale from dark blue to red, as densities range from about $3.0 \times 10^{-15}$ to $4.8 \times 10^{-11}$ g cm$^{-3}$, respectively. The scale saturates to white at even higher densities. The burst phase occurs at about 6 ORP. By about 10 ORP, the disk has transitioned into its asymptotic behavior.

### 3.2.1 Overall Evolution

Table 1 summarizes some simulation results. Here, "Duration" refers to the simulation length measured in ORPs. Other entries are defined as they are introduced in the text.

As shown in Figure 2 for Irr 15K, these disks go through the same evolutionary phases as described in Paper II and Paper III. The initial *axisymmetric cooling* phase ends with a *burst* phase of rapid growth in several discrete global nonaxisymmetric spiral modes that rapidly redistrbute mass. The first noticeable effect of irradiation is a delay in the start of the burst phase, with later onset for higher $T_{irr}$, as indicated by the $t_1$ entries in Table 1. Heating by irradiation slows the approach to a strongly unstable $Q$-value. As in Paper III and Cai et al. (2006), these disks do not form persistent protoplanetary clumps, in contrast to Boss (2001, 2006). Instead, after the burst, the disks become more axisymmetric again during a *transition* phase and then settle into a quasi-steady *asymptotic* phase of nonaxisymmetric GI activity, where heating and cooling are in rough balance and where average quantities change slowly (see also Lodato & Rice 2005). The approximate start of the asymptotic phase is indicated by $t_2$ in Table 1 and does not appear to be delayed as $T_{irr}$ increases. In this paper, we focus on the disk behavior during the asymptotic phase, because the bursts may be an artifact of our initial conditions (see Papers II and III).

Figure 3 compares midplane densities at $t = 14$ ORP for all four cases. Not surprisingly, the densest spirals appear in the No-Irr case. For Irr 50K, even though the disk has not yet had a chance to settle into an asymptotic state at the time shown, the spiral structure already appears much weaker than in the Irr 25K disk. Following Cai et al. (2006), to quantify differences in GI strength, we compute Fourier amplitudes $A_m$ for the nonaxisymmetric structure over the whole disk during the asymptotic phase as follows (see also Imamura et al. 2000):

$$A_m = \frac{\int \rho_m r dr dz}{\int \rho_0 r dr dz}$$

(5)

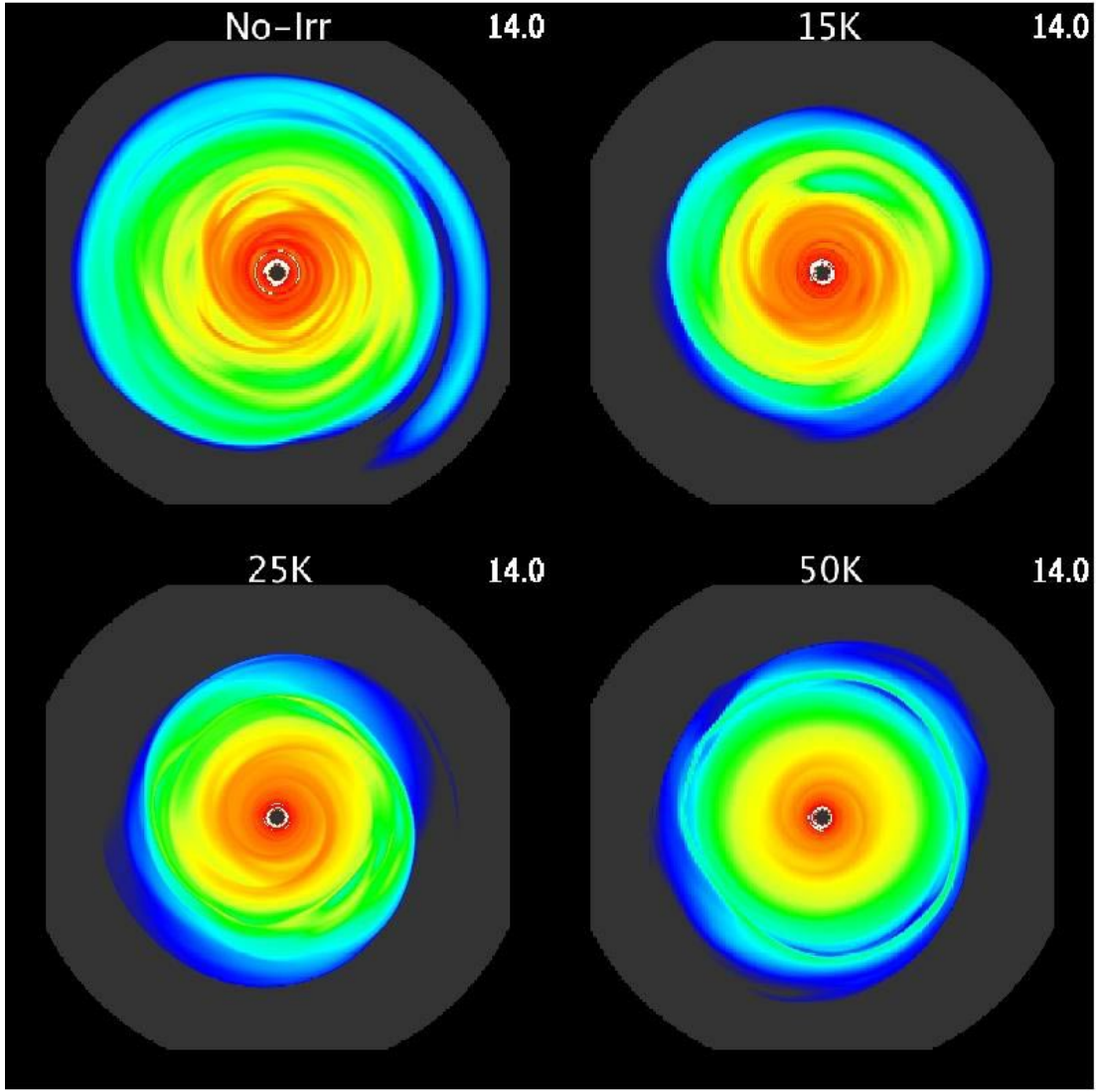

**Figure 3.** Midplane density maps at 14 ORP for the four cases labelled. Each square enclosing the disk is 155 AU on a side. The color scale is the same as in Fig. 2.

where $\rho_0$ is the axisymmetric component of density and $\rho_m$ is the amplitude of the $m$th Fourier component for a decomposition of the density in azimuthal angle $\varphi$.

As a measure of the total nonaxisymmetry, $A$, a sum of the $A_m$ over all resolved $m \geq 2$, is computed. The $A_1$ is excluded because our star is fixed at the origin, and so $m = 1$ may not be treated realistically (see Paper III). This sum is quite variable with time, and so the entry $\langle A \rangle$ in Table 1 is $A$ averaged over the 13.0-14.0 ORP time period. $\langle A \rangle$ is greatest for No-Irr and decreases with increasing $T_{irr}$, in agreement with what we see in Figure 3. Overall, stronger envelope irradiation leads to lower nonaxisymmetric amplitudes. For Irr 50K, not only is its $<A>$ value the smallest, but its GI amplitudes continue to decrease after this time interval. For example, when $<A>$ is re-evaluated for 14.3 - 15.3 ORPs, it is only $\sim 0.28$. The stabilization of the disk by $T_{irr} = 50$K is corroborated by the high $Q$ value at 14.9 ORPs (see Table 1 & Fig. 5). A crude fit to $A(t)$ yields an exponential damping time for the GIs of $\sim 5$ ORP.

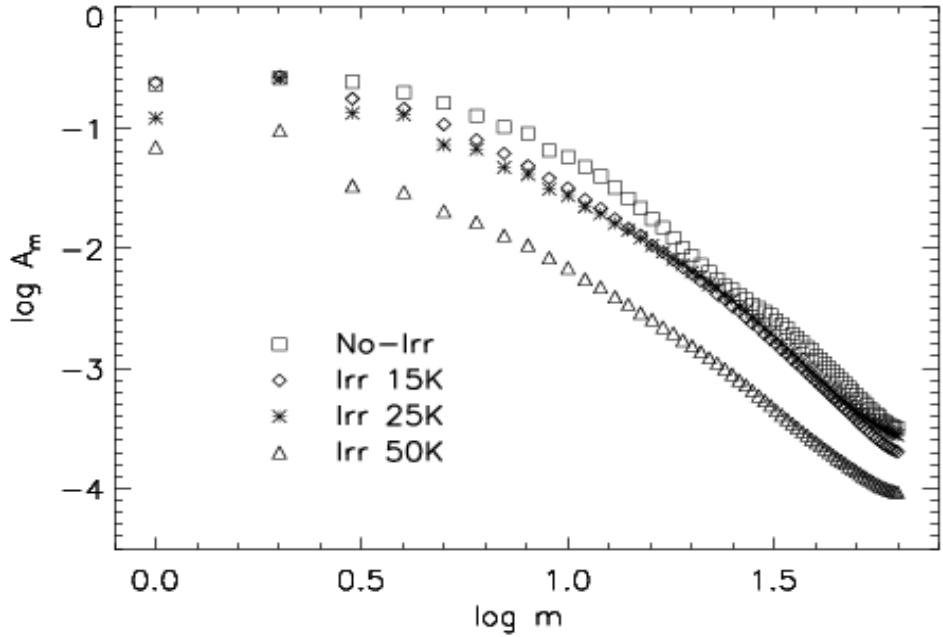

**Figure 4**. $<A_m>$ distribution over the last two ORPs of the simulations. As the irradiation increases, the higher-order modes are preferentially damped, and $m = 2$ becomes more pronounced in the profile. Moreover, the profile tends toward a shallower drop off at large $m$ as the low-order modes are suppressed.

To show the relative strengths of the nonaxisymmetric Fourier components, a distribution of $<A_m>$ averaged over the last two ORPs for each simulation is presented in

Figure 4*. A higher $T_{irr}$ progressively suppresses high-order components. In contrast, moderate irradiation seems to preserve $m = 2$ selectively, and it is the last component of nonaxisymmetric structure to fade away. For large $m$, the $A_m$ are roughly fit by an $m^{-3}$ profile for No-Irr, Irr 15K, and Irr 25K and by a shallower $m^{-2.7}$ for Irr 50K. As in Paper III, the GIs appear to be dominated by global low-order modes, with superposed gravitoturbulence (Gammie 2001) manifested by the higher $m$-values. This interpretation agrees with Fromang et al. (2004)'s hydrodynamics simulations where low-order modes tend to dominate the nonlinear spectrum of GIs. With a high level of irradiation ($T_{irr} = 50K$), all GI modes are damped.

The Toomre $Q = c_s \kappa / \pi G \Sigma$, where $c_s$ is the adiabatic sound speed and $\kappa$ is the epicyclic frequency, is calculated using azimuthally-averaged values, which are localized to the midplane for the variables $c_s$ and $\kappa$. The evolution of $Q$ with time is discussed in detail in Cai (2006); here we focus on late times. Figure 5 shows midplane $Q(r)$ profiles at $t = 14.9$ ORP during the asymptotic phase, and Table 1 lists the $Q_{min}$ and the radial range $\Delta r_{Q<2}$ over which $Q < 2$ at this late time. Clearly, the low-$Q$ region becomes narrower and the $Q_{min}$ becomes higher for higher $T_{irr}$. Because it is these low-$Q$ regions that drive the GIs, this is another manifestation of GI suppression by irradiation. The radial width of the low-$Q$ region differs among the runs and so does the mass contained within it. To characterize the difference more precisely, mass-weighted average $Q$-values for $t = 15$ ORP are computed between 15-35 AU and are given in Table 1 as $<Q_{late}>$. The values of $<Q_{late}>$ in Table 1 and the curves in Figure 5 show that $Q$ generally increases as $T_{irr}$ increases. In particular, as clearly shown in Figure 5, $Q$ for Irr 50K is nowhere below 2.0, which indicates that the Irr 50K disk is stable against GIs (Pickett et al.1998, 2000; Nelson et al. 2000). The GIs developed earlier in this disk when $T_{irr} = 25$ K are being damped, as confirmed by the decay of $<A>$ mentioned earlier.

------

* The $m = 1$ component seen in Fig. 4 is probably generated by nonlinear mode coupling (Laughlin & Korchagin 1996), as evidenced by its phase incoherence in a periodogram analysis.

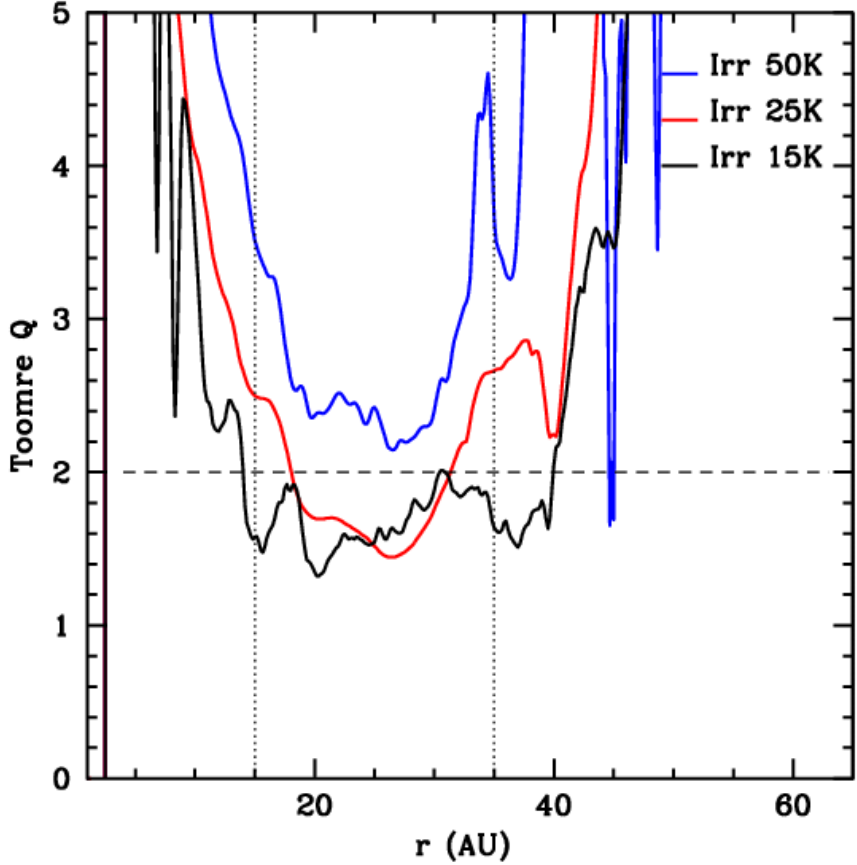

**Figure 5.** The Toomre $Q(r)$ at 14.9 ORPs for the three irradiated disks. The dotted lines delineate the radial range over which $<Q_{late}>$ is computed for Table 1, and the dashed line is $Q = 2$.

This behavior led us to wonder whether the disk would have become unstable at all if we set $T_{irr} = 50$K at $t = 0$. However, when we change the midplane temperatures $T_{mid}$ to 50K wherever $T_{mid} < 50K$ in the initial disk, the initial $Q_{min}$ only rises to about 1.69, probably not enough to prevent instability during the initial cooling phase. The difference in $Q_{min}$ values between the initial disk at 50K and the asymptotic disk at 50K is caused by the decrease in surface density due to radial mass transport in the evolved disk (see §3.2.3). A similar test in which we reset the initial $T_{mid}$ to 70K yields a $Q_{min}$ of 2.0. Indeed, in a low azimuthal resolution test run with $T_{irr} = 70K$ that covers 2.8 ORP, $Q_{min}$ saturates at about 2.2, well above the marginally stability limit for nonaxisymmetric GIs of about 1.5 to 1.7 (see Durisen et a. 2007). This demonstrates that, for our disk, there is a critical value of $T_{irr}$ between 50 and 70K, call it $T_{crit}$, where a simulation started at $t = 0$ with $T_{irr} > T_{crit}$ would never become unstable.

### 3.2.2 Energetics

Figure 6 shows total internal energy and total energy losses due to cooling as functions of time. The "optically thick" cooling curves are total energy losses in the interior of the disk due to diffusion as determined from the divergence of the flux; the "optically thin" cooling curves are the net energy losses based on integrating equation (1). As in Figure 2 of Cai et al. (2006), due to our limited vertical resolution and the Eddington fit near the photosphere (see §2.2), the thick curves effectively include energy losses in most of the photospheric layers for most columns through the disk. The thin curves tally additional cooling from extended layers above the photospheric cells, plus the parts of the outer disk that are optically thin all the way to the midplane (hereafter, the "outer thin disk"). Figure 6 shows that, while the total energy loss in Irr 15K is dominated by cooling in the optically thick region, the optically thick cooling in Irr 25K is essentially zero. The net cooling that occurs is primarily due to optically thin emission, especially after about 8 ORP, even though a greater radial range of the disk midplane is optically thick for higher $T_{irr}$.

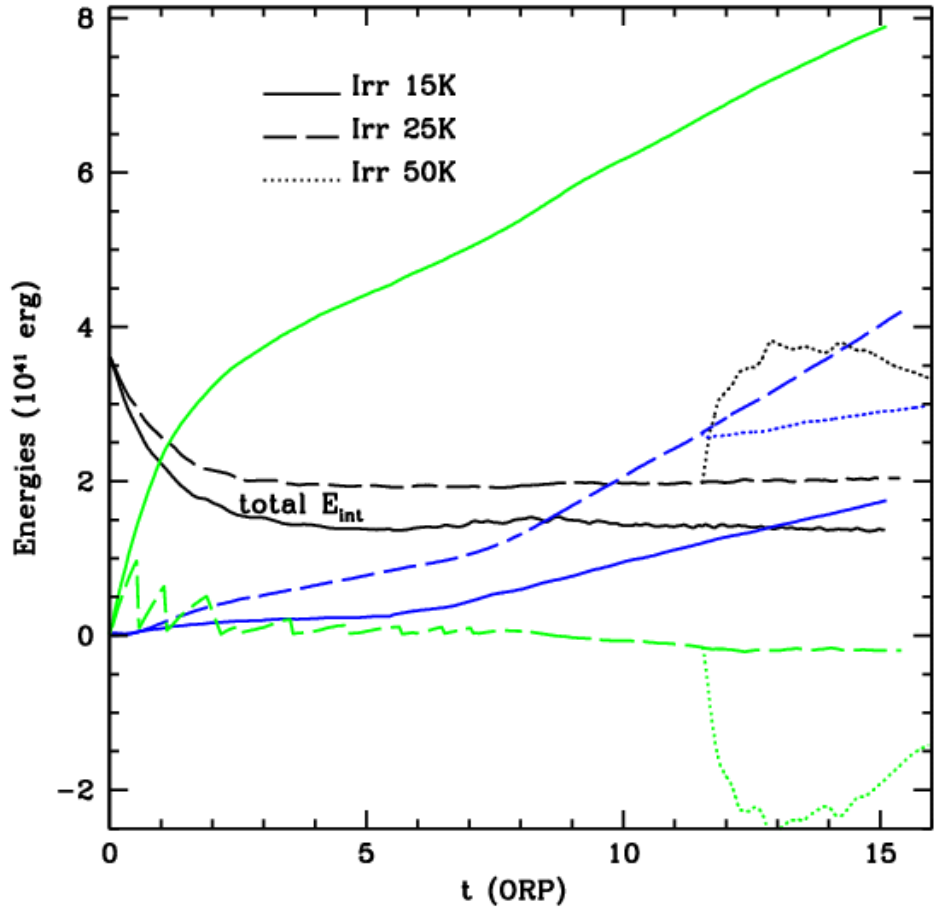

**Figure 6.** Cumulative total energy loss as a function of time due to radiative cooling in optically thick (green curves) and optically thin (blue curves) regions. The two black curves that start at $3.6 \times 10^{41}$ ergs and lie in the middle of the diagram and the dotted tail rising upward at $t = 11.5$ ORP for 50K show how the total disk internal energies (labelled $E_{int}$) change with time for the three calculations. The type of the line denotes which run it is plotting, as shown in the figure legend.

When $T_{irr}$ is increased to 50K at $t$ = 11.5 ORP, the optically thick disk heats rapidly until about $t$ = 13 ORP, after which it cools, as revealed by the positive slope of the $E_{int}$ curve and the negative value of the optically thick "cooling" curve for Irr 50K when the curves begin at $t$ = 11.5 ORP. The Irr 15K and Irr 25K internal energy curves indicate relaxation to a state of quasi-steady balance of heating and cooling, while Irr 50K is still undergoing transient adjustments.

As in Cai et al. (2006), the final global net cooling time ($t_{cool}$ in Table 1) is obtained by dividing the final total internal energy by the final total net radiative energy loss rate using the slopes of the thin and thick curves in Figure 6. Table 1 shows that $t_{cool}$ is longer for a disk with higher $T_{irr}$. When $T_{irr}$ = 50K, it reaches ~10 ORP. However, it should be noted that, especially for Irr 25K and Irr 50K, instantaneous cooling times computed for individual vertical columns of the disk fluctuate spatially and temporally by factors of several and can even change sign. As reported in Paper III for the No-Irr case, the overall and local cooling times for irradiated disks also tend to increase as they evolve. Gammie (2001) showed that disks fragment only when $t_{cool}$ < about a rotation period. The $t_{cool}$ values in Table 1 suggest that none of our disks should fragment, consistent with what we observe.

We can estimate the actual envelope irradiation being put into the disk from summing $\sigma T_{irr}^{4}(1-\exp[-\tau_{mid}])$ times the top area of each vertical column over the entire disk surface, where $\tau_{mid}$ is based on the Planck mean opacity (see §2.2). For simplicity, the additional radial irradiation in the Irr 15K case is not counted. The result is tabulated in Table 1 as $L_{env}$(code). In all three cases, the numbers are close to the analytic estimations of $L_{env}$ (Table 1, row 12) with a disk radius $R_d$ = 50 AU (see §3.2.0). For comparison, the total *net* cooling "luminosity" $L_{disk}$ (Table 1, row 14) is computed by adding the total *net* energy loss rates at the end of each simulation. As $T_{irr}$ increases, $L_{env}$ becomes $>> L_{disk}$. So, the absorption and re-emission of the envelope energy input for $T_{irr}$ = 25 and 50K far outweighs the net cooling of the disk that counterbalances heating from GI activity.

Figure 7 compares the vertical density and temperature structures in the asymptotic phase for Irr 25K (top) and No-Irr (bottom). Envelope irradiation makes the disk physically thicker and generates a nearly isothermal structure with temperature $T$ close to $T_{irr}$. The temperature increase over No-Irr in the outer disk is dramatic. Superficially, this appears to justify the Boss thermal bath approach (see §2.3), but it is still not clear that Boss obtains the correct BC for the optically thick regions of his disks.

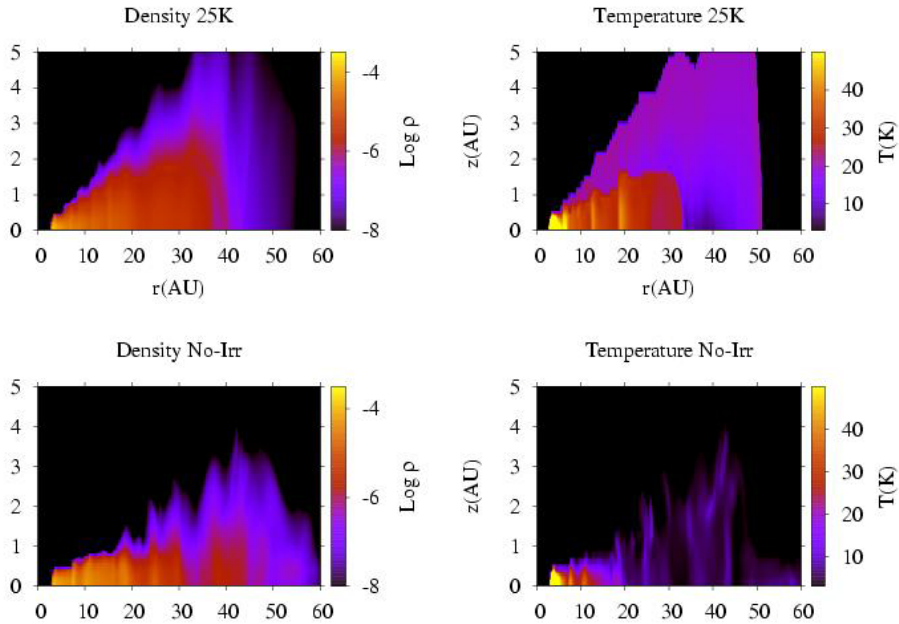

**Figure 7.** Meridional cross-sections of density (left) and temperature (right) at $t$ = 12.0 ORP (3000 yr). The Irr 25K case is at the top and No-Irr case at the bottom. The density and temperature scales are displayed on the right of each panel, ranging from $\log(\rho/\mathrm{gcm}^{-3})$ = -8 to -3 and T = 3 to 50 K, respectively. Note that the vertical scale is exaggerated for clarity; the boxes span 60 AU radially and 5 AU vertically.

Our result also agrees qualitatively with D'Alessio et al. (1997).

The low temperature pocket near the midplane in the optically thin outer region of the Irr 25K disk is probably an artifact of using mean opacities evaluated at different temperatures in equation (1) and of incomplete radiative coupling of cells in the optically thin regions (see §2.2). For $a_{max} = 1\mu m$ and $T < 100K$, the Planck mean is greater than Rosseland mean (Appendix A of Paper III). As a result, the envelope irradiation gets completely absorbed before reaching the midplane in some columns for the outer disk even though the Rosseland $\tau_{mid} < 2/3$. A related problem is that, in columns where the envelope irradiation is not drained by the midplane, the surviving photons should continue through and be at least partially absorbed on the other side of the disk. These upward moving photons are not included in our scheme. The existence of the overcool region near the midplane suggests that not many columns are affected by this omission. These two problems do not change our main conclusions, because they both tend to bias our calculations toward *underestimating* the effect of irradiation.

### 3.2.3 Mass Transport and Redistribution

Strong mass transport occurs during the burst phase, though it becomes weaker as $T_{irr}$ increase. Here we limit our discussion to the asymptotic phase, because the initial bursts may be an artifact of our initial conditions (see Paper III). Mass transport rates for this phase averaged over the last 3 ORPs are plotted in Figure 8, and Table 1 (row 15) gives a spatial average of the inward mass transport rates computed between 9 AU and the radius at which the radial mass flow shows a major sign change in Figure 8, which we call the "mass inflow/outflow boundary". For the irradiated disks in Figure 8, both the peak and average $\dot{M}$'s decrease as $T_{irr}$ increases from 15K to 25K. However, the $\dot{M}$'s in both these irradiated disks appear to be substantially *higher* than for No-Irr. This is perhaps the most surprising result of our simulations. As will be discussed at greater length in §3.2.4, it appears that *mild* irradiation *enhances* the gravitational torques and hence the mass transport by selectively suppressing higher-order spiral structure while leaving the dominant global two-armed ($m = 2$) modes relatively intact (see also Figure 3). The mass transport for the Irr 50K run is not shown in Figure 8, because the GIs are damping and so the mass transport observed is a dying transient that should not be compared to asymptotic phase results.

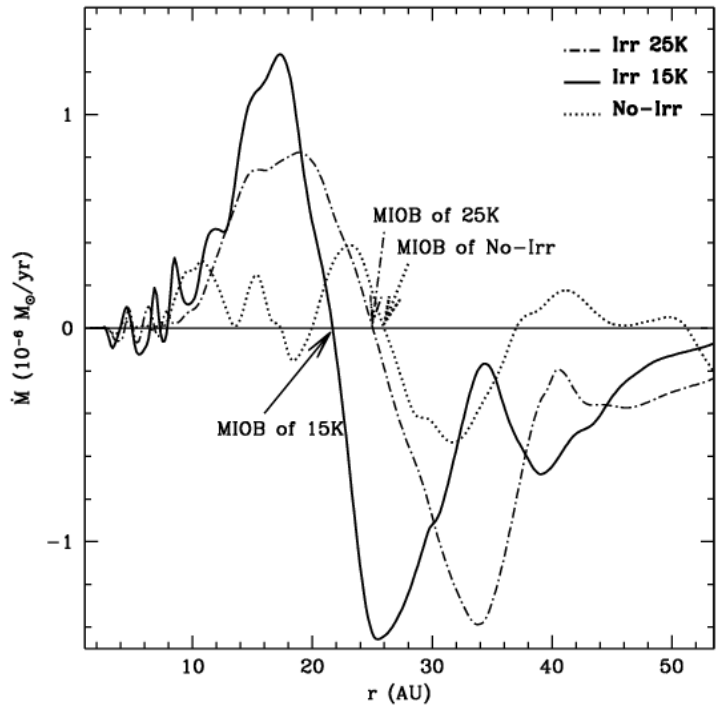

**Figure 8.** The mass inflow rate calculated by differences in the total mass fraction as a function of radius. The time interval is taken to be the last 3 ORP for each calculation. The mass inflow/outflow boundary (MIOB) is labeled for each curve. In the No-Irr case, we adopt 26AU as the major MIOB because, over a long period of time, the designated MIOB is the typical sign change boundary.

During the burst phase, the initial surface density profile $\Sigma(r) \sim r^{-0.5}$ is destroyed. As in Paper III, during the asymptotic phase, the inner disks show substantial concentrations of mass in annular regions, while the azimuthally averaged $\Sigma(r)$ in the outer disks can be reasonably well fit by a power law. For Irr 15K, a fit of $\Sigma \sim r^{-p}$ gives $p \sim 2.85$ between 22 and 40 AU. The Irr 25K disk is similar but with $p \sim 3.5$ between 25 and 40 AU. Compared with No-Irr (see Paper III for details), the irradiated disks fall off more steeply outside 40 AU, the initial outer radius of the disk. Because it is GI torques during the burst that expand the disk, the steeper fall off probably results from weaker bursts for irradiated disks. In Paper III, we found that $\Sigma(r)$ for the No-Irr in the asymptotic phase could be fit over a very wide range of radii by a Gaussian. This is not the case for the irradiated disks, and so we do not present Gaussian fits.

Like the simulations presented in Papers II and III and in Cai et al. (2006), dense gas rings are produced in the inner portions of the irradiated disks and appear to still be growing when the calculations end (see Cai 2006 for more detail). Nearly axisymmetric rings form at about 6.5 and 8 AU for Irr 15K and are obvious in azimuthally-averaged surface density plots of inner disks, as shown in Figure 9. The surface density maxima at 11, 15, and 20 AU represent strong radial concentrations of mass, which exhibit strong nonaxisymmetric behavior and are not rings. Durisen et al. (2005) suggested that rings and radial concentrations may act as regions of accelerated planetary core formation (see also Rice et al. 2004, 2006). For $r > 10$ AU, the physics in our simulations are well resolved. There, the radial concentrations are real and associated with mass transport by discrete global two-armed spirals. We have begun higher resolution simulations, to be reported elsewhere, to test whether or not the innermost axisymmetric rings are physically real or are artifacts of poor vertical resolution in the inner disk. Regardless, the inner regions of real disks should be kept hot and GI-stable by physical processes not included in our code (see, e.g., Najita et al. 2006), and rings are expected to form at boundaries between GI active and inactive regions, as shown already in Paper I, Paper II, and Durisen et al. (2005).

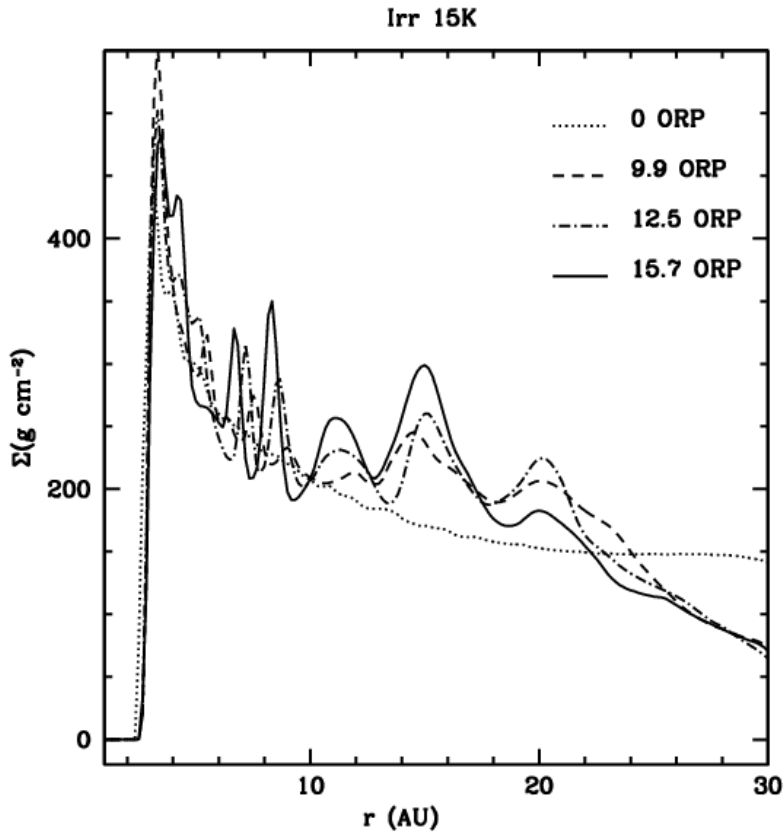

**Figure 9.** Azimuthally-averaged surface density distribution for the inner disk of Irr 15K. Shown are the averaged surface density profiles at 0, 9.9, 12.5, and 15.7 ORP.

### 3.2.4 Torques, Modes, and Effective $\alpha$

In Paper III, the gravitational and Reynolds (or hydrodynamic) torques associated with mass and angular momentum transport were calculated for the No-Irr disk. Similar calculations for the irradiated disks are shown in Figure 10. We only plot the gravitational torques, because the Reynolds torques are poorly determined and tend to be considerably smaller. In both the Irr 15K and Irr 25K cases, the mass inflow/outflow boundaries shown in Figure 8 roughly align with the global maxima of the gravitational torques (Figure 10, upper left panel) when averaged over the same interval as the mass transport rates, consistent with what is observed in the No-Irr disk (Paper III). A slight degree of *mis*alignment is probably due to the different sampling rate used for the various time-averaged values and the difficulty of calculating accurate Reynolds stresses in our code (see Paper III).

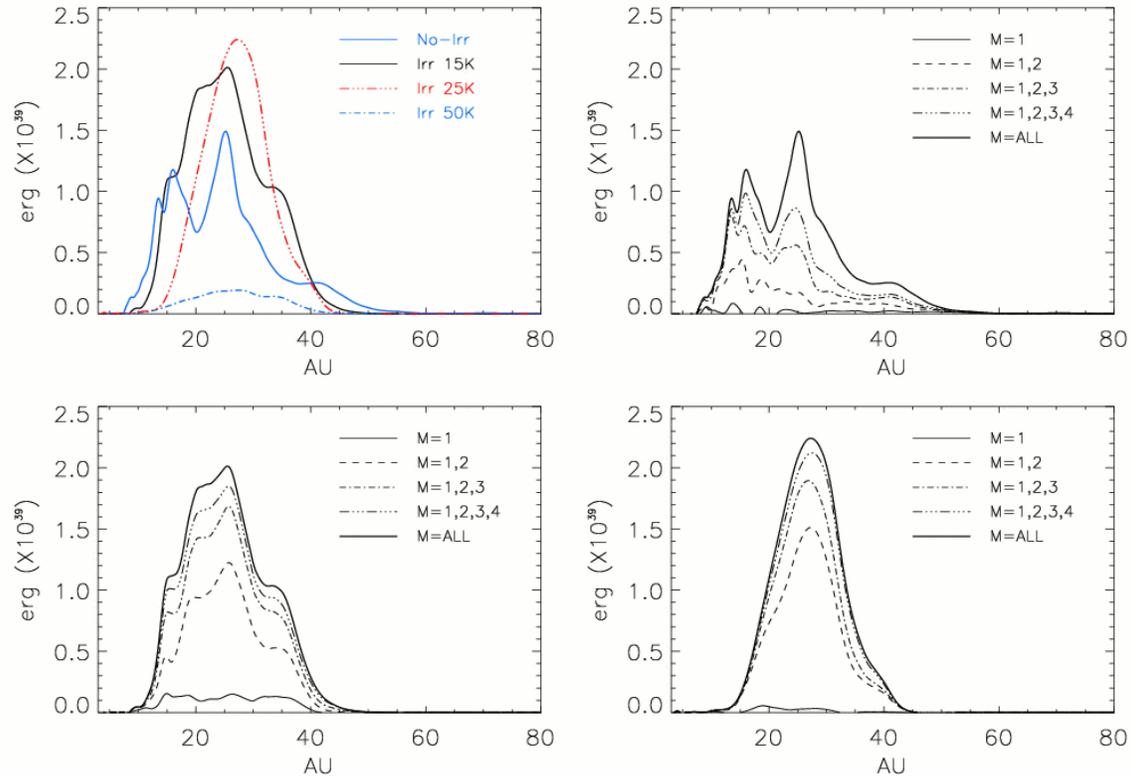

**Figure 10.** *Upper left.* The gravitational torque for each simulation time-averaged over the last three ORPs. The torque becomes smoother and larger for moderate irradiation, while strong irradiation suppresses the torque. *Upper right.* A Fourier deconstruction of the gravitational torque for the No-Irr simulation over the same time period. The curve labeled M = 1 is the torque due only to the Fourier component $m = 1$, the curve labeled M = 1,2 is the torque due to $m = 1$ and the torque due to $m = 2$ added together, and so on. *Lower left.* The same for the Irr 15K simulation. *Lower right.* The same for the Irr 25K simulation. As $T_{irr}$ increases, the $m = 2$ component becomes more dominant in the torque. In all of the irradiated cases, almost all the torque can be accounted for by the four lowest-order modes.

The *m*-wise deconstructions in the upper right and the two lower panels of Figure 10 demonstrate that, as $T_{irr}$ increases, the $m = 2$ or two-armed component of the nonaxisymmetric structure becomes more dominant. This is consistent with the GI spectra of Figure 4 and suggests that $m = 2$ spirals are the principal mode for mass and angular momentum transport in the irradiated disks. A periodogram analysis of $m = 2$ for Irr 15K (as in Figure 11 of Paper III for No-Irr) confirms that several discrete coherent global pattern periods have corotation radii that line up with the peaks in the torque profile in the lower left panel of Figure 10. The upper left panel of Figure 10 also shows that the magnitudes for the torques in Irr 15K and Irr 25K are comparable and, at their peaks, are roughly 1.4 times the maximum torque of No-Irr. This is consistent with the high average mass transport rates in these runs reported in §3.2.3 and Table 1. We speculate that the $m = 2$ spirals, being more global, are more efficient at producing net torques, while high-order modes tend to produce more fluctuations and cancellation on smaller scales, as evidenced by the multi-peaked character of the No-Irr curves in Figure 8 and Figure 10. So, the torque increases when mild envelope irradiation preferentially suppresses modes with $m > 2$. On the other hand, as shown for the Irr 50K case in Figures 5 and 10 (upper left), too much irradiation suppresses GIs and hence mass transport altogether by heating the disk to stability.

Also, as in Paper III, we calculate an effective Shakura & Sunyaev (1973) $\alpha$ from the torque profile, as shown in Figure 11. Despite the larger torques and greater domination by $m = 2$ for $T_{irr} = 15$ and 25K, the effective $\alpha$ is lowered as the irradiation is increased. This may at first seem paradoxical. However, the effective $\alpha$ due to gravitational torques is proportional to the vertically integrated gravitational stress tensor divided by the square of the sound speed times the disk surface density (see equation [20] of Paper III). Effectively, $\alpha$ is a dimensionless measure of the relative strength of gravitational stresses and gas pressure. In the asymptotic phase of the Irr 15K and Irr 25K simulations, the sound speed is higher due to the increase in the midplane temperature caused by irradiation, and $\Sigma$ is also higher in the 15 to 35 AU region because the initial GI burst is weaker. We have verified numerically that the increase in $\Sigma c_s^2$ does outweigh the higher torques as $T_{irr}$ increases to give lower $\alpha$ values. It is not clear from these simulations what would happen to $\alpha$ if we increased $T_{irr}$ for an asymptotic disk with the same initial $\Sigma$. This will be explored in future papers. Figure 11 also shows that the uniformity of the irradiation tends to drive $\alpha$ toward a constant value over a large fraction of the GI-active region. The GIs in Irr 50K are damping, and so $\alpha$ should approach zero in this case on the same time scale ($\sim$ 5 ORP). The $\alpha$ values at 30 AU for No-Irr, Irr 15K, and Irr 25K are $2.5 \times 10^{-2}$, $1.3 \times 10^{-2}$, and $8 \times 10^{-3}$, respectively, comparable to values reported in other GI studies for global simulations with similar cooling times (Lodato & Rice 2004, Paper III, Boley et al. 2007c).

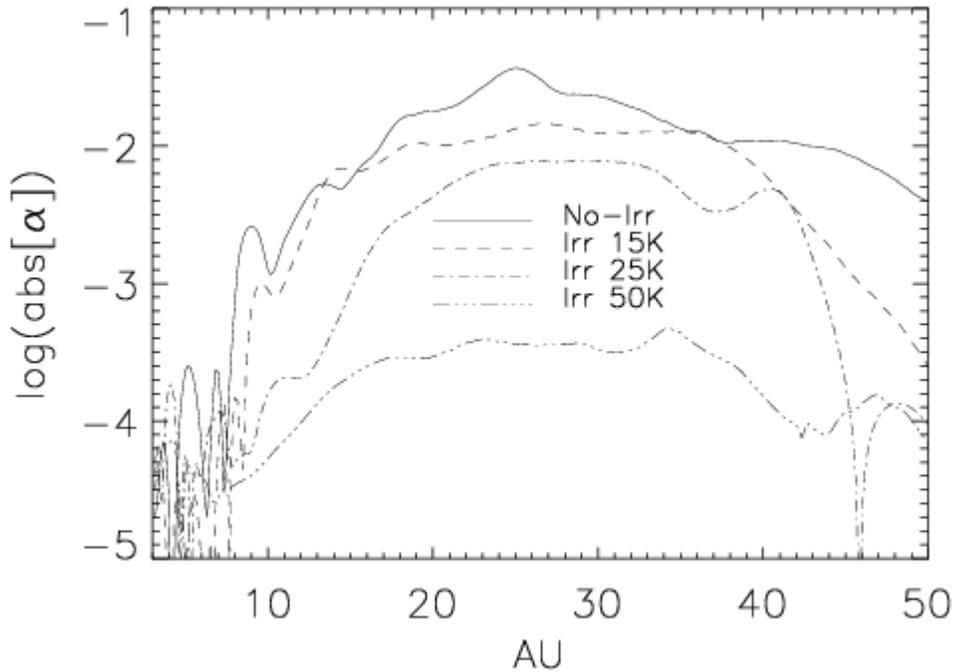

**Figure 11.** Effective $\alpha$ averaged over the last 3 ORP of each simulation as a function of radius. Despite the increase in total torque for moderate irradiation, the effective $\alpha$ is suppressed as $T_{irr}$ increases due to an increase of the midplane temperature.

### 3.3 Massive Embedded Disk

Recently, a few embedded disks have been resolved and studied in detail by millimeter-wave interferometry. Examples include L1551 IRS 5 (Osorio et al. 2003), SVS 13 (Anglada et al. 2004) and IRAS 16293-2422B (Rodriguez et al. 2005). These YSOs are all very young (Class 0 or I), with the stars and massive circumstellar disks embedded in huge envelopes. Although massive, these disks are small in size. Except for SVS 13, the Toomre $Q$-values are estimated to be small in the outer part of the disks. For example, Osorio et al. (2003) fit the SEDs with a composite model that includes almost all components of the L1551 IRS5 binary system, including the envelope, the circumbinary disk, and both circumstellar disks. The disk parameters were derived (see Table 2 of Osorio et al. 2003) based on fitting the observed fluxes at (mostly) millimeter (mm) wavelengths with irradiated steady-state α-disk models. They find that the disk masses in L1551 IRS 5 are comparable to those of the accreting stars, with a Toomre $Q < 1$ in the outer disks. Using the radiative scheme described in §2.2, we have performed a disk simulation tuned to their derived parameters. In fact, our group has modeled such massive, young, stubby disks in earlier papers (e.g., Pickett et al. 1997, Pickett et al. 1998).

### 3.3.1 Initial Model

Following the procedures outlined in Paper I, an initial axisymmetric equilibrium model was generated with a specified $M_d/M_{tot}$, $R_d/R_p$, and power-law index $p$ of the surface density distribution. We choose $M_d/M_{tot} = 0.4$ to resemble the Northern disk of L1551 IRS 5, where $M_{star} = 0.3\ M_\odot$, and $M_d = 0.2\ M_\odot$ (see Table 2 of Osorio et al. 2003). The outer radius of the disk $R_d$ is set to 15 AU, where 1 AU = 16 cells, and $p = 1.0$. After creating an equilibrium star/disk model, a modified 2D code was used to remove the star and replace it with a point mass potential. The disk model is then evolved in 2D for a few thousand steps to quench the waves generated by removing the star. The ORP is now defined as the orbital period at 14.5 AU or about 76.6 years. The initial grid has (256, 128, 32) cells in $(r, \varphi, z)$ above the midplane. The initial $Q$ profile shown in Figure 12 has a $Q_{min}$ of about 0.9, a little higher than Osorio et al. (2003)'s prediction for the Northern disk but still below 1. This difference owes partly to the use of the isothermal sound speed in $Q$ by Osorio et al., in contrast to the adiabatic sound speed ($\gamma = 5/3$) we use. Following Osorio et al., an envelope irradiation temperature $T_{irr} = 120K$ is applied. The maximum grain size $a_{max}$ in the D'Alessio et al. (2001) opacity is set to be 200 μm to be similar to the Northern disk, and the mean molecular weight is 2.34 proton masses.

### 3.3.2 The Evolution

As in the simulations of §3.2, the disk remains fairly axisymmetric for the first 3 ORPs of evolution, which is surprisingly long for a disk with initial $Q_{min} < 1$. Toomre's (1964) stability analysis is based on a thin disk approximation, and so the finite thickness of this disk introduces a correction factor to $Q$ when used as a stability criterion (Romeo 1992, Mayer et al. 2004). According to Mayer et al. (2004), the correct $Q^*$ is related to the conventional $Q$ by $Q^* = (1+2\pi h_d/\lambda_{mu})Q$. The quantity $\lambda_{mu}$ is the "most unstable wavelength" and equals $0.55\lambda_{crit}$ (Binney & Tremaine 1987), where the "Toomre wavelength" $\lambda_{crit} = 4\pi^2 G\Sigma/\kappa^2$. The quantity $h_d$ is the disk scale height, which we compute as $h_d = 0.5\Sigma/\rho_{mid}$, following Romeo (1992), where $\rho_{mid}$ is the midplane density. The correction factor is about 1.5 to 2 in the outer disk, and the uncorrected initial $Q_{min}$ of 0.9 becomes $Q^*_{min} \approx 1.4$, as shown in Figure 12. This value indicates that, instead of being violently unstable to axisymmetric disturbances, this disk is only marginally unstable to nonaxisymmetric modes, similar to the initial model presented in §3.1.

By $t = 3.5$ ORP, spiral structure is apparent in the disk. Then the arms expand outward until 4.6 ORP (see Figure 13). During this burst phase, the disk is dominated by global $m = 2$ and 3 spiral modes and expands greatly in both the radial and vertical directions. Not only was the number of radial cells doubled to 512, but the number of vertical cells had to be quadrupled to 128 in order to avoid losing too much mass off the grid. Unfortunately, the calculation had to be stopped at $t = 4.6$ ORP due to a numerical difficulty that arises when the dense spiral arms become separated by very low-density inter-arm regions. The flux-limited diffusion calculation in the $r$ and $\varphi$-directions apparently breaks down due to the high resulting density and temperature contrasts and produces unphysical results. Figure 12 shows the $Q(r)$ and $Q^*(r)$ profiles at $t = 0$ and 3.6 ORP. By $t = 3.6$ ORP, both $Q$s increased, primarily in the outer disk. Shock heating is an important heating source, as it is in the burst phase of other simulations (Papers II and III). The intense envelope irradiation does not seem to affect the disk evolution, at least at this stage.

The actual envelope irradiation shining onto the disk, calculated as in §3.2, is about $7.7 \times 10^{33}$ erg/s at $t = 4$ ORP, whereas the net $L_{disk}$ is almost 5 orders of magnitude lower, which indicates a very long overall net cooling time. In fact the optical depth to the midplane at $r \sim 10$ AU is $\sim 4 \times 10^3$. A large optical depth and slow radiative cooling are expected for the high surface density and large grain size.

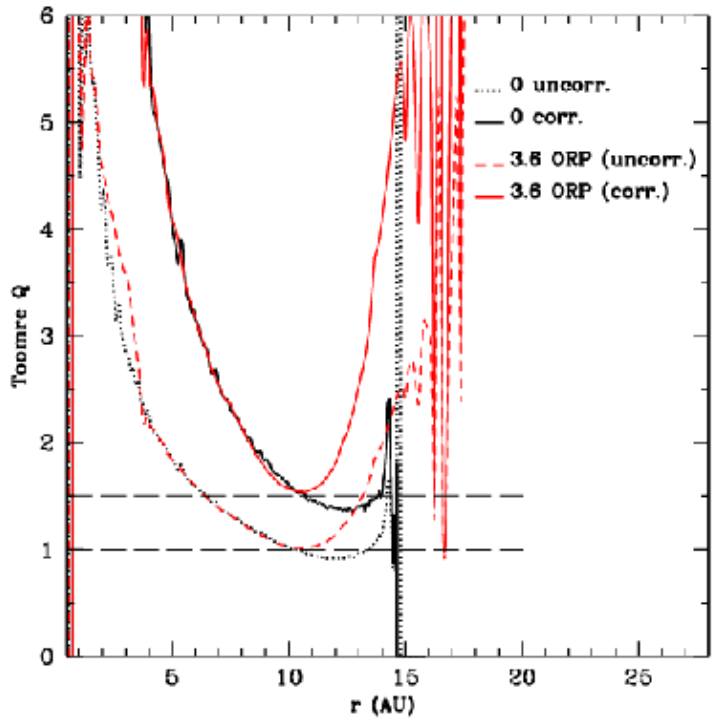

**Figure 12**. Q(r) profile at t = 0 and 3.6 ORP. The black curves represent the initial Q profile, with dotted curves denoting the uncorrected and solid curves being corrected Q. The red curves are Q(r) at 3.6 ORP, where the dashed curve shows the uncorrected Q, and the solid one is the corrected Q. The values Q = 1 and 1.5 are plotted in black dashed lines for reference.

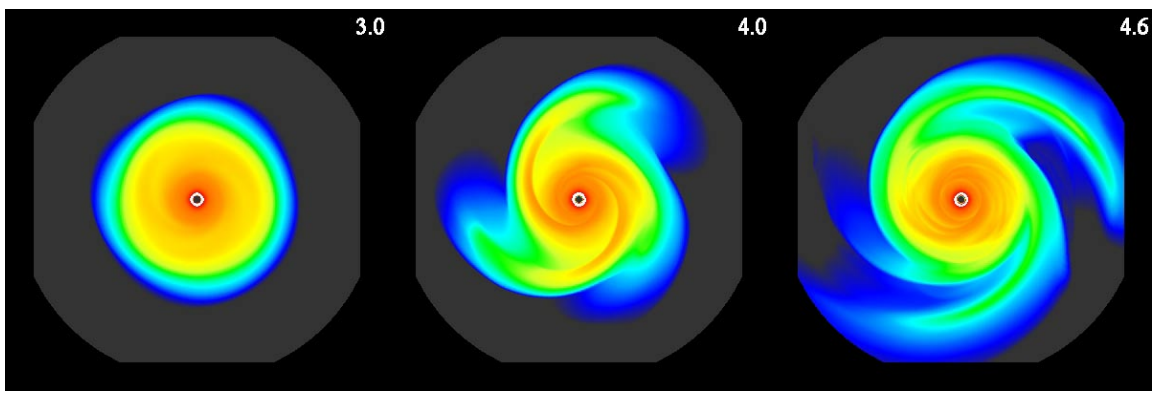

**Figure 13**. Selected midplane density contours for the L1551 IRS 5-like disk. Each square enclosing the disk is 58 AU on a side. The number at the upper right corner of each panel denotes the time elapsed in ORP units. Densities are displayed on a logarithmic color scale from dark blue to red, with densities

ranging from about $4.6\times10^{-14}$ to $2.3\times10^{-9}$ g cm$^{-3}$, except that the scale saturates to white at even higher densities.

### 3.4 Simulations with a Boss-like Boundary Condition

To investigate whether the apparent disagreement between our results and those of Boss is primarily due to radiative boundary conditions, two "Boss-like" calculations were carried out with the Boss-like radiative BC (see §2.3). To compare with the irradiated disk simulations presented in §3.2, the background temperature $T_B$ is set to 15K or 25K, with the same initial axisymmetric model (§3.1). We refer to the resulting simulations as Boss 15K and Boss 25K, respectively. Following Boss (2001, 2002), the artificial bulk viscosity (AV) is turned off.

Both Boss-like disks develop "bursts" after only a few ORPs, then gradually adjust and appear to approach a quasi-steady state, similar to the irradiated disks. The disk structures do not resemble those of Boss disks, and no clumps are produced. A comparison of the bursts between Irr 25K and Boss 25K indicates that the presence of AV heating tends to maintain higher temperatures and lower the GI amplitudes. In the Boss 25K case, most of the disk cools down to 25K by 3 ORPs.

In both Boss-like calculations, by $t = 8$ ORP, just before the disks enter the asymptotic phase, the inner disk becomes very thin and dense and the inner edge fragments, but with an unphysical-looking pattern. We suspect the fragmentation is artificial in nature due to the high density and low resolution there (Truelove et al. 1997, Nelson 2006). A calculation of the local Jeans length and the local grid resolution reveals that the Jeans number is about 1/8 to 1/10, smaller than the upper limit (1/4) that Truelove et al. (1997) suggested for stability. However, a proper Truelove et al. criterion may be code specific, because it is likely to depend on grid geometry, advection scheme, and differencing scheme. Since we use a cylindrical grid while Truelove et al.'s work was based on Cartesian grid simulations, we suspect that the fragmentation is numerical and that the critical Jeans number for artificial fragmentation is code dependent. Nelson (2006) pointed out that, for a disk, a Toomre criterion is more appropriate than the Jeans analysis. Nevertheless, according to Nelson's equation (16), the Toomre wavelength is comparable or even larger than the local Jeans wavelength in our case ($Q > 1$). Another possible contributing factor to the numerical instability is the breakdown of our radiative routine in the inner disk due to insufficient vertical resolution (Nelson 2006). To investigate all this, a stretch of the Boss 15K was re-run, with doubled resolution in all three directions. The result is that the inner edge did *not* fragment, at least for the 0.7 ORP of the rerun, which confirms the artificial nature of the fragmentation. In fact, in Boss's own disk simulations with spherical coordinates, his Jeans number is not far from 1/4, especially at the loci of dense clumps (Boss 2002).

Our Boss-like disks seem to reach higher nonaxiymmetric amplitudes. However, without data in the asymptotic phase, there is not much that can be concluded from the comparisons, except that the Boss-like BCs seem to allow more rapid cooling and contraction of the inner disk. As mentioned in §2.3, there are some BCs that can only be implemented with a similar starting model. The use of an initial disk similar to the one used in the Boss simulations seems necessary. We constructed such an initial model according to the formulations in Boss (1993, 2003) and carried out two disk simulations using both a Boss-like BC and our radiative scheme. Drastic differences were observed. We are trying to understand the results in a collaborative effort with Boss. More details will be included in a forthcoming paper.

## 4. DISCUSSION

### 4.1 The Effect of Irradiation on Gravitational Instabilities

In this paper, we have reported a series of simulations for a disk with a radius of 40 AU and a mass of 0.07 $M_\odot$ around a young star of 0.5 $M_\odot$ to study how variations in the amount of IR envelope irradiation onto the disk affects gravitational instabilities. As the irradiation temperature $T_{irr}$ is increased, the nonaxisymmetric amplitudes of GIs decrease, the Toomre $Q$-values increase, the energy input from irradiation becomes more dominant in the disk thermal equilibrium, the effective $\alpha$ derived from GI torques decreases, and the overall net cooling times become longer. In particular, in the Irr 50K simulation, when $T_{irr}$ is increased from 25 to 50K during the asymptotic phase, the GIs damp out on a time scale comparable to the cooling time. The increase in $Q$-values when $T_{irr}$ is increased indicates that the disk becomes gravitationally stable for this $T_{irr} = 50$K case (see Fig. 5). Thus, we conclude that mild envelope IR irradiation weakens GIs, and strong irradiation suppresses them altogether. In all cases, envelope irradiation raises the outer disk temperature significantly above its initial values, as found by D'Alessio et al. (1997).

Perhaps our most interesting and unexpected result is how the radial mass inflow rate $\dot{M}$ varies with $T_{irr}$. In simulations with idealized cooling, where the cooling time is held fixed everywhere, Paper II found that the asymptotic mass transport rate is inversely proportional to cooling time. Here we find that, with mild irradiation ($T_{irr}$ = 15K), the $\dot{M}$ becomes significantly *larger* than for no irradiation at all, even though the overall cooling time is longer (see Table 1). As discussed in §§3.2.1 and 3.2.4, mild irradiation ($T_{irr}$ = 15 and 25K) selectively suppresses high-order spiral structure without affecting the strength of the two-armed ($m$ = 2) modes that dominate the global torques and drive the mass transport. We think the reason behind this selective suppression of the high-order modes is that irradiation increases the sound speed ($c_s$), which tends to stabilize the instabilities with short wavelengths (Toomre 1964). This increase in $c_s$ is reflected by the increase of the asymptotic $Q$-values in the irradiated disks as $T_{irr}$ is increased (cf. Table 1 and Fig.5). Mild irradiation thus increases the total torque, but, as explained in §3.2.4, the effective $\alpha$ decreases somewhat as $T_{irr}$ increases due to the larger values of $\Sigma c_s^2$ in the asymptotic states of irradiated disks. The torque increase associated with isolation of the $m$ = 2 waves reinforces arguments in Papers II and III that GIs are an intrinsically global phenomenon (see also Balbus & Papaloizou 1999). In fact, the same mode-dependent suppression is observed in simulations with varied metallicity and grain size presented in Cai et al. (2006), though no discernable difference in mass transport rates was noticed in that paper, probably due to the low accuracy used in the analysis. As irradiation is strengthened slightly from $T_{irr}$ = 15 to 25K, the mass inflow rate decreases somewhat but is still higher than for no irradiation at all. At $T_{irr}$ = 50K, irradiation is sufficient to suppress GIs entirely in the asymptotic phase and mass inflow shuts off.

While $T_{irr}$ = 50K is enough to damp GIs in the asymptotic phase, we found in §3.2.1 that this is not a high enough $T_{irr}$ to prevent GIs from growing in the initial disk.

So, there seem to be three regimes of behaviors with increasing $T_{irr}$: 1) For $T_{irr} < T_{crit1}$, the disk is able to become gravitationally unstable and remains unstable thereafter. 2) For $T_{crit2} > T_{irr} > T_{crit1}$, the disk becomes unstable, but the GIs are eventually damped. 3) For $T_{irr} > T_{crit2}$, the disk never becomes unstable. For our disk, we find that $T_{crit1}$ is between 25 and 50K and $T_{crit2}$ between 50 and 70K. The difference between $T_{crit1}$ and $T_{crit2}$ is due to the drastic redistribution of mass during the burst.

We suspect that our main conclusion – that irradiation tends to weaken and even suppress GIs – probably generalizes qualitatively to other types of irradiation, such as stellar irradiation and external irradiation from nearby stars. On the other hand, our preliminary calculation of a highly embedded massive disk (§3.3) demonstrates that a disk can become unstable even with $T_{irr}$ as high as 120K. Weakening and stabilization of GIs by irradiation makes direct planet formation by GIs in irradiated disks less likely, and realistic modeling of the GIs in such embedded disks must take envelope irradiation into account.

## 4.2 Comparison with Analytic Arguments

Rafikov (2005) has argued analytically that, for realistic opacities, an unstable disk and fast enough cooling for fragmentation are incompatible constraints in the planet-forming regions of radiatively cooled disks. More recently, in agreement with our own results in Paper III, he has furthered argued (Rafikov 2006) that, contrary to Boss (2004), convection cannot substantially change this conclusion. The simulations in this paper show that irradiation makes this problem worse. Modifying Rafikov's arguments by using $Q_0 = 1.5$ instead of 1 as the stability threshold for GIs (as also critiqued by Boss 2005), we can evaluate $\Sigma_{inf}$ and $T_{inf}$ in Rafikov's (2005) equations (7) and (10)) based on our disk parameters at $r = 20$ AU, where $Q$ is low (see Fig. 5). With his $f(\tau)$ factors, we find that Rafikov's (2005) fragmentation conditions (6) and (9) predict no fragmentation, consistent with what we see in the asymptotic phase of the simulations. In the asymptotic phase, we expect our disks to hover near marginal stability. With $Q_0 = 1.5$, Rafikov's (2005) criterion (2) for GIs predicts midplane densities that are within a factor of two of what we see at 20 AU for the Irr 15K and Irr 25K simulations, which is a good agreement for such an approximate analysis.

By examining the stability of steady-state accretion disk models, Matzner & Levin (2005) also argue analytically that gravitational fragmentation will be suppressed under realistic conditions and is not a viable planet formation mechanism. Their analysis considered both the local instability and a global one, where the latter applies to very early phases when disks are massive and might be subject to global SLING instability (see Shu et al. 1990). Their treatment of disk irradiation is qualitatively similar to what our irradiated disks achieve in the asymptotic phase, and their estimate for the fraction of stellar surface flux tapped ($f_F$, equivalent to $L_{env}/L_{star}$ in §3.2) puts their typical disk irradiation intensity between our Irr 25K and Irr 50K. Not surprisingly, we reached similar general conclusions.

However, both Rafikov and Matzner & Levin's local GI analysis are based on the Gammie fragmentation criterion (2001), which may not hold precisely for a disk with realistic radiative cooling (Johnson & Gammie 2003). Moreover, the critical value in Rafikov's (2005) criterion (3) changes with the first adiabatic index (Rice et al. 2005), which will vary in real disks (see, e.g., Boley et al. 2007a).

## 4.3 Comparisons with Other Numerical Work

Boss (2002) explored effects on GIs of varying some of his thermodynamic assumptions, including variations in the "outer disk temperature" $T_0$, which controlled the initial $Q_{min}$. Because Boss, in these tests, resets $T_0$ every time the disk cools below this temperature, $T_0$ is analogous to our $T_{irr}$ in terms of its effect (see Fig. 7). In one model with $T_0 = 150$K, the initial $Q_{min}$ reached 2.6, and "only weak spiral arms'' developed. This high $T_0$ model resembles our Irr 50K disk, where intense envelope irradiation damps GIs gradually. Boss (2002) also noted that clump formation is slightly delayed in time in models with moderate $T_0$ when compared to those with lower $T_0$, similar to the delayed burst we report in §3.2.1. So, indirectly, our main finding that irradiation tends to weaken GIs was anticipated in the Boss (2002). A critical difference, however, is that our disks do not fragment into dense clumps, regardless of $T_{irr}$. Boss (2004) attributes fragmentation in his disks with radiative cooling to rapid cooling by convection, whereas our cooling times in Table 1 are too long for fragmentation to occur, and, as we will now explain, we do not see and do not expect convection in these irradiated simulations.

The superadiabatic regions seen during the axisymmetric phase of No-Irr (Paper III) are, for the most part, absent in Irr 15K and Irr 25K. There is some convection during roughly the beginning of each irradiation simulation, probably due to our isentropic starting condition, but it disappears after about $t = 2$ ORP, well before the GI burst. Because No-Irr convects until the onset of GIs, we surmise that the irradiation makes the temperature profile too shallow for superadiabatic regions to form even without GI-activity. There are some superadiabatic regions at the interface between the interior and the atmosphere of the disk due to the sudden temperature drop shown in Figure 1 (see Paper III for more details), but because the optical depths are low at the drop, convection is not present at these altitudes. During the asymptotic phase, we find no sign of convection even though a large volume is still optically thick in both irradiated disks. So, even the mild irradiation of the $T_{irr}$ = 15K case is sufficient to suppress convection. For the irradiated simulations, there are strong vertical motions, but they are associated with shock bores, as in No-Irr (Boley & Durisen 2006, Paper III). We strongly suspect that the vertical motions seen by Boss (2004) are not convection. Even if convection does occur in a disk, we would not expect it to produce substantially more rapid cooling, because all disk energy transported vertically must still ultimately be radiated away (Paper III, Ravikov 2006).

Another question of interest to GI researchers is whether there is time in a disk's evolution during which GIs can be approximated by a local description (Laughlin & Rozyczka 1996, Balbus & Papaloizou 1999, Gammie 2001, Lodato & Rice 2004, Paper II, Paper III). As summarized in §4.1, all the calculations presented in this paper strengthen the case made by Paper III that GIs are dominated by *global* low-order modes. Here, we find that mild envelope irradiation enhances global behavior by selectively suppressing high-order modes, while not reducing the amplitudes of the lowest-order modes, especially $m$ = 2. On the other hand, the effective $\alpha$ due to the GIs becomes more uniform in $r$ as $T_{irr}$ increases, and it may be possible to crudely approximate the mass transport by GIs with an $\alpha$ prescription. However, the picture of mass slowly diffusing inward is misleading, because global low-order modes induce large fluctuations in the radial motions of fluid elements and lead to shock bores (Boley & Durisen 2006). Moreover, substructure caused by resonances with the global modes will likely be missed in a local description. Such substructures may be important to planet formation (Durisen et al. 2005) because solids are marshalled by gas drag into regions of pressure maxima (Weidenschilling 1977; Haghighipour & Boss 2003; Rice et al. 2004).

## 4.4 Implications for Real Protoplanetary Disks and Planet Formation

Real disks are probably irradiated on the surface not only by the envelope but also by a variety of other sources, such as the central star and nearby stars, as evidenced by recent molecular line observations (e.g., Najita et al. 2003, Qi et al. 2005). Although different forms of irradiation require different treatments, our simple blackbody approximation gives us some general insight. All forms of surface irradiation will cause heat to flow into the deeper layers of the disk. Judging by the luminosities in Table 1, we surmise that, if we start with a marginally unstable disk, GIs become completely suppressed once the irradiation energy input becomes more than one or two orders of magnitude greater than the energy dissipation rate in the disk interior that can be sustained by GIs. Such energy input rates by external radiation are not unusual (D'Alessio et al. 2001). At levels below this, irradiation significantly alters the structure of GIs, tending to make them more ordered and global, with notable enhancement of gravitational torques and mass inflow rates. The selective damping of higher order modes means that significant irradiation makes fragmentation of a disk much less likely.

Our treatment of envelope irradiation assumes that $T_{irr}$ does not vary with radius or time. Irradiation of real disks, especially by their central stars, is probably spatially inhomogeneous and time variable. Variability raises the interesting possibility that the nature of GIs could vary significantly over GI-active regions of the disk in strength, mass transport rate, and structure. In a disk simulation where we adopt a variable $T_{irr}$ profile $T_{irr}(r) \sim r^{-0.5}$, with a total $L_{env}$ matching that of Irr 25K run, preliminary results, to be reported elsewhere, suggest that stellar irradiation may be somewhat more effective in suppressing GIs than envelope irradiation. Protracted shadowing or strong illumination of disk regions could have a profound impact on GI behaviour as well. The complete suppression of GIs in the Irr 50K simulation requires about one global cooling time, or about two thousand years. This is probably the characteristic response time to temporal variations in irradiation for the 10s AU regions of young disks for the opacities we assume. In the future, we need to model envelope irradiation in more detail, by calculating the reprocessing of the stellar irradiation, as in Matzner & Levin (2005). We also need to attempt realistic treatment of other forms of irradiation.

For an embedded disk, the envelope not only acts as a hot blanket, but it also replenishes mass onto the disk continuously through infall, which may make the disk more unstable to GIs (Mayer et al. 2004, Vorobyov & Basu 2005, 2006). On the other hand, this process will also result in an accretion shock (Banerjee, Pudritz & Holmes 2004, Banerjee & Pudritz 2006) that might tend to suppress GIs. As shown by Chick &

Cassen (1997), efficient accretion could significantly increase both the disk and envelope temperatures. Our Irr 50K case shows that such heating could damp GIs completely. Taken together, the fate of an embedded disk may depend on which of the physical processes (envelope irradiation, accretion shock heating, and mass accumulation) is most efficient. There are many factors that could affect disk fragmentation in the embedded phase, so it may still be a bit too early to say that GIs cannot produce protoplanetary clumps in an embedded disk under realistic conditions.

While tending to inhibit fragmentation and hence the direct formation of planets by disk instability, irradiation tends to enhance the global character of GIs. As a result, irradiation still allows the production of rings and radial concentrations of gas, and so a hybrid theory remains viable, where GIs accelerate core accretion by providing dense structures into which solids can become concentrated (Rice et al. 2004, 2006, Durisen et al. 2005).

Some attempts were made in Paper III to generate spectral energy distributions (SEDs) from the simulations. It is premature, however, to do so in our case, because we do not include a detailed model for the envelope, which would contribute significantly to the SEDs (e.g., Calvet et al. 1994, Miroshnichenko et al. 1999). Another limitation is our use of a grain size distribution similar to that of the interstellar grains, which has proven to be inappropriate for protoplanetary disks (D'Alessio et al. 2001), even for those in the embedded phase (Osorio et al. 2003). We use small grains here in order to make direct comparisons with the No-Irr simulation from Paper III.

When we simulate a small, massive disk with radiative cooling (§3.3), the disk goes unstable even with an envelope irradiation of 120K and larger grain sizes. Unless the disk parameters for L1551 estimated by Osorio et al. (2003) do have large systematic errors (especially the disk masses), this implies that the development of GIs and strong spiral structure is inevitable in heavily embedded phases, when the disks are comparable in mass to their central stars. Powerful millimeter instruments are being built which hold promise for the detection of such spiral structures (see, e.g.,Wolf & D'Angelo 2005). Unfortunately, our L1551-like disk simulation could not be integrated more than a few outer orbit periods, and so we cannot say with any confidence how this disk will behave over longer periods of time. It is not even clear that such a massive disk will settle into an asymptotic phase. It may instead be subject to episodic bursts (Lodato & Rice 2005, Vorobyov & Basu 2005, 2006). What we can say with some confidence is that the cooling times are very long, and the disk is unlikely to fragment.

During our short integration of the L1551-like disk, the radial mass distribution undergoes considerable modification. The disk expands to a size approaching the Roche lobe radius of the system in a few orbit periods. In this sense, the Osorio et al. disk parameters are not physically self-consistent, i.e., the low-$Q$ disk they describe is strongly and globally unstable. Because the binary separation is only ~ 40AU, about three times the radius of either disk (see Table 2 of Osorio et al. 2003), any further modeling of GIs in an L1551-like disk would have to take the gravitational effect of the binary companion into account. In fact, tidal truncation has been suggested as an explanation for the small disk radii (Artymowicz & Lubow 1994, Rodriguez et al. 1998, Lim & Takakuwa 2005).

Inasmuch as the L1551-like disk is an extremely massive, optically thick disk, it is worthwhile to consider what other accretion mechanisms may be viable. Energetic particles have a typical attenuation of 100 g cm$^{-2}$ (Stepinski 1992), and any nonthermally ionized layer would likely be fairly shallow, but a reasonable estimate for its depth is beyond the scope of this paper. As argued by Hartmann et al. (2006), a thin MRI-active layer likely will be unable to create Reynolds stresses that affect the midplane of the disk as seen in layered accretion simulations with thick (>10%) MRI-active layers (Fleming & Stone 2003). Moreover, the temperatures only become high enough to ionize thermally alkalis (T>1000K; e.g., Gammie 1996) in approximately the first AU of the simulated disk. However, it should also be noted that T>1000K may be insufficient to trigger a thermal MRI because dust will deplete ionized species. It may be necessary for temperatures to approach dust sublimation before a thermal MRI becomes effective (e.g, Desch 2004). It seems that the primary heating mechanisms for the interior of this disk are related to gravity, e.g., gravitational contraction in the vertical direction, GIs, and tidal forcing by a binary companion. Because GIs are dominated by global modes (Pickett et al. 2003; Mejia et al. 2005; Boley et al. 2006), treating the disk as an alpha disk (e.g., Osorio et al. 2003) is likely to misrepresent the evolution of the system.

Even though our models are directly applicable only to low mass stars, our main conclusions may well apply to Herbig Ae disks, given their similarity to T Tauri disks, as found by FUSE (e.g., Grady et al. 2006). However, it is probably inappropriate to extend the scaling to more massive star/disk systems like Herbig Be stars. Recent observations (e.g., Monnier at al. 2005, Mottram at al. 2007) suggest distinctly different accretion scenarios between the two groups.

Finally, our results indicate that, regardless of whether net cooling is dominated by optically thick or thin regions, cooling times are too long to permit fragmentation. Convection is suppressed during all phases of evolution of the irradiated disks (except the L1551-like disk), and, even if convection did occur in the irradiated cases as it did in the axisymmetric phase of No-Irr, it is not expected to lower the cooling times enough for fragmentation. All energy must ultimately be radiated away (Paper III, Rafikov 2006). Moreover, we are confident that our code allows for convection under the appropriate conditions, inasmuch as we have tested it against analytic problems (see Boley et al. 2007c).

## 5. CONCLUSION

We have presented 3D radiative hydrodynamics simulations of disks around young low-mass stars demonstrating that infrared irradiation by a surrounding envelope significantly affects the occurrence and behavior of gravitational instabilities. By selectively weakening high-order structure, mild irradiation enhances the global character of the resulting spiral waves and increases mass inflow rate. On the other hand, strong irradiation suppresses gravitational instabilities entirely. In all cases, irradiation tends to inhibit disk fragmentation and protoplanetary clump formation. Future simulations require more detailed treatment of the envelope structure, reprocessing of starlight, and the effects of accretion onto the disk from the envelope. Improvements in the gas equation of state (Boley et al. 2007a), better algorithms for radiative transfer (Boley et al. 2007c), consideration of other forms of irradiation, and inclusion of the effects of binary companions (e.g., Nelson 2000, Mayer et al. 2005, Boss 2006) are also needed. Work by our group along these various lines is planned or underway.

*Acknowledgments.* This work was supported in part by NASA Origins of Solar Systems grants NAG5-11964 and NNG05GN11G and Planetary Geology and Geophysics grant NAG5-10262. We thank S. Basu, A.P. Boss, N. Calvet, C.F. Gammie, L. Hartmann, R.E. Pudritz, and R. Rafikov for useful comments and discussions related to this work, and P. D'Alessio for providing mean molecular weight and dust opacity tables. We especially would like to thank the anonymous referee whose comments led to substantial improvements in the presentation of our results. K.C. and A.C.B. acknowledge the generous support of a CITA National Postdoctoral Fellowship and a NASA Graduate Student Researchers Fellowship, respectively. This work was supported in part by systems obtained by Indiana University by Shared University Research grants through IBM, Inc. to Indiana University.